 \documentclass[12pt,preprint]{aastex}

\newcommand{\xmmnewton}{{\it XMM-Newton}}

\newcommand{\einstein}{{\it Einstein}}
\newcommand{\chandra}{{\it Chandra}}

\newcommand{\ctb}{CTB~109}

\newcommand{\nh}{\mbox {$N_{\rm H}$}}
\newcommand{\nhtwo}{\mbox {$N_\mathrm{H_2}$}}
\newcommand{\htwo}{H$_{2}$}

\newcommand{\about}{$\sim$\kern.03em}

\shorttitle{Shocked Molecular Gas in SNR CTB~109}
\shortauthors{Sasaki et al.}

\begin{document}


\title{Evidence for Shocked Molecular Gas in the \\ Galactic SNR CTB~109 (G109.1--1.0)}


\author{Manami Sasaki\altaffilmark{1},
Roland Kothes\altaffilmark{2,3},
Paul P. Plucinsky\altaffilmark{1},
Terrance J. Gaetz\altaffilmark{1},
Christopher M. Brunt\altaffilmark{4}}

\altaffiltext{1}{Harvard-Smithsonian Center for Astrophysics,
    60 Garden Street, Cambridge, MA 02138; msasaki@cfa.harvard.edu,
    pplucinsky@cfa.harvard.edu, tgaetz@cfa.harvard.edu.}
\altaffiltext{2}{National Research Council of Canada, Herzberg Institute of
Astrophysics, Dominion Radio Astrophysical Observatory, P.O. Box 248,
Penticton, BC V2A 6J9, Canada; roland.kothes@nrc-cnrc.gc.ca.}
\altaffiltext{3}{Department of Physics and Astronomy, University of Calgary,
2500 University Drive NW, Calgary, AB T2N IN4, Canada.}
\altaffiltext{4}{School of Physics, University of Exeter,
Stocker Road, Exeter EX4 4QL, United Kingdom,
brunt@astro.ex.ac.uk.}


\begin{abstract}
We report the detection of molecular clouds around the X-ray bright interior 
feature in the Galactic supernova remnant (SNR) \ctb\ (G109.1--1.0). 
This feature, called the Lobe, has been previously suggested to be the result 
of an interaction of the SNR shock wave with a molecular cloud complex.
We present new high resolution X-ray data from the \chandra\ X-ray 
Observatory and new high resolution CO data from the Five College
Radio Observatory which show the interaction region with the cloud
complex in greater detail.
The CO data reveal three clouds around the Lobe in the velocity interval 
$-57 < v < -52~\mathrm{km}~\mathrm{s}^{-1}$. 
The velocity profiles of $^{12}$CO at various parts of
the east cloud are well fit with a Gaussian; however, at the position where
the CO cloud and the Lobe overlap, the velocity profile has an additional 
component towards higher negative velocities. The molecular hydrogen density
in this part of the cloud is relatively high 
(\nhtwo\ $\approx 1.9 \times 10^{20}$~cm$^{-2}$), whereas the foreground 
absorption in X-rays (\nh\ $\approx 4.5 \times 10^{21}$~cm$^{-2}$), obtained
from \chandra\ data, is lower than in other parts of the cloud and in the
north and south cloud.
These results indicate that this cloud has been hit by the SNR blast wave on 
the western side, 
forming 
the bright X-ray Lobe.
\end{abstract}




\section{Introduction}

The progenitor stars of core-collapse supernova explosions form in
giant molecular clouds. Since these massive stars have a short lifetime
many of them end their lives while the parental clouds are still nearby
and may even still harbor small star forming regions that produce stars
of lower mass. According to \citet{1999A&A...351..459C} in galaxies like ours 
about 70\% of all supernova explosions are of type II and should explode close 
to the dense 
clouds from which they were formed. After these stars explode, strong shocks 
are driven into the clouds, heating, compressing, dissociating, and 
accelerating the gas leading to a large variety of observable effects.
A picture book example is the Galactic supernova remnant (SNR) IC~443 on which 
most studies of SNR-molecular cloud interactions have been focused
\citep[][and references therein]{1998ApJ...505..286S,2000A&A...362L..29B,2002ApJ...572..897K}. 
But recently more and more SNRs have been discovered interacting with
molecular clouds, e.g.\ W28, W44, 3C~391 
\citep[e.g.,][and references therein]{2000ApJ...544..843R,2003ApJ...585..319Y},
and many others, among them the Galactic SNR \ctb\ (G109.1--1.0).

\ctb\ was first discovered as an SNR in X-rays with \einstein\ 
\citep{1980Natur.287..805G} and in radio in the 610~MHz Galactic plane survey 
\citep{1981ApJ...246L.127H}. It has a
semi-circular morphology in both the X-ray and the radio and is
located next to a giant molecular cloud (GMC) complex in the west.
This semi-circular morphology 
suggests that the SNR shock has been stopped entirely by the GMC complex,
and that the appearance is not simply due to absorption.
A linear feature in CO (`CO arm') extends from the GMC complex to the local 
X-ray minimum in the northern half \citep{1987A&A...184..279T} implying
that a part of the GMC complex extends in front of the remnant (see
Fig.\,\ref{chandraco}). The cold interstellar medium in which the remnant
is embedded has been studied in detail by \citet{2002ApJ...576..169K}.
The most puzzling X-ray morphological feature in \ctb\ is the bright, extended 
interior region known as the `Lobe'. The X-ray spectrum from the Lobe obtained 
with \xmmnewton\ is completely thermal \citep{2004ApJ...617..322S}. 
The Lobe could be the result of a hole in the GMC allowing the X-ray
emission through with little or no absorption or it could be the
result of intrinsically brighter emission due to an interaction
between the shock and the cloud. In order to investigate the later
hypothesis we obtained new high resolution X-ray and CO data.

\section{Observations}

\subsection{CO data}

Observations of the $^{12}$CO and $^{13}$CO (J=$1-0$) spectral lines, 
at 45\arcsec\ resolution, were obtained using the Five College Radio Astronomy
Observatory (FCRAO) 14~m antenna in March 2003. The telescope was equipped
with the 32~element SEQUOIA focal plane array \citep{eric99}.
The data were acquired through on-the-fly mapping, in which the telescope
was scanned continuously across the sky while reading out the
spectrometers at regular intervals of 11\farcs25.
Calibration to the $T_{A}^{*}$ scale was done using the chopper
wheel method \citep{ku81}, and the data were converted to
the radiation temperature scale ($T_{R}^{*}$) by correcting for
forward scattering and spillover losses ($\eta_\mathit{fss}$~=~0.7).
The 1024-channel spectrometers were set to a total bandwidth of
25~MHz ($\sim65~\mathrm{km}~\mathrm{s}^{-1}$) centered on 
$-45~\mathrm{km}~\mathrm{s}^{-1}$.
Following recording of the data, the spectra were
converted onto a regular grid of 22\farcs5~pixel spacing using the
FCRAO {\it otftool} software.

The new data have higher sensitivity than the CGPS data 
\citep{2002ApJ...576..169K} and are fully Nyquist sampled. 
The higher sensitivity and the full sampling allow us to
detect the faint clouds around the Lobe and study them in 
great detail.

\subsection{\chandra\ data}

The \xmmnewton\ observations have shown that the X-ray bright Lobe is thermal
and seem to indicate an interaction between the shock wave and a molecular
cloud \citep{2004ApJ...617..322S}. Therefore, we proposed an additional deep
observation with \chandra\ to probe the shock-cloud interaction region at 
higher angular resolution. The observation 
was performed using the Advanced CCD Imaging Spectrometer (ACIS) in
full-frame,
timed-event mode with an exposure of 80~ksec (ObsID 4626). The data were taken 
in the energy band of \about0.3 -- 10.0~keV. The ACIS-I array covered the 
northeast part of the SNR and the northern tip of the Lobe was observed at the 
aimpoint. The data are analyzed with CIAO 3.2.2 and CALDB 3.1.0.
The complete analysis of these data including a detailed spectral analysis 
of the whole area will be presented in a different paper. 

Here, we present the high-resolution X-ray image of 
the Lobe of
 \ctb\ obtained 
with \chandra\ ACIS-I. The image is binned with a size of 4 pixels (1 pixel =
0.492\arcsec) and smoothed with a Gaussian with a sigma of 2 pixels (the
original pixels binned by 4). 
X-ray spectra which are extracted at regions corresponding to CO clouds are 
also discussed, in order to obtain the absorbing foreground hydrogen column 
density (\nh). The spectra are binned with a minimum of 50 counts per 
bin and analyzed using the X-ray spectral analysis tool XSPEC. To fit the
spectra, we use a model for a thermal plasma in non-equilibrium ionization 
with variable abundances (VNEI) and hydrogen column density, \nh, for the 
foreground absorption (PHABS). 

\section{Discussion}

\subsection{CO, Infrared, and X-ray Data}

\begin{figure}
\centering
\caption{\label{chandraco}
{\bf See 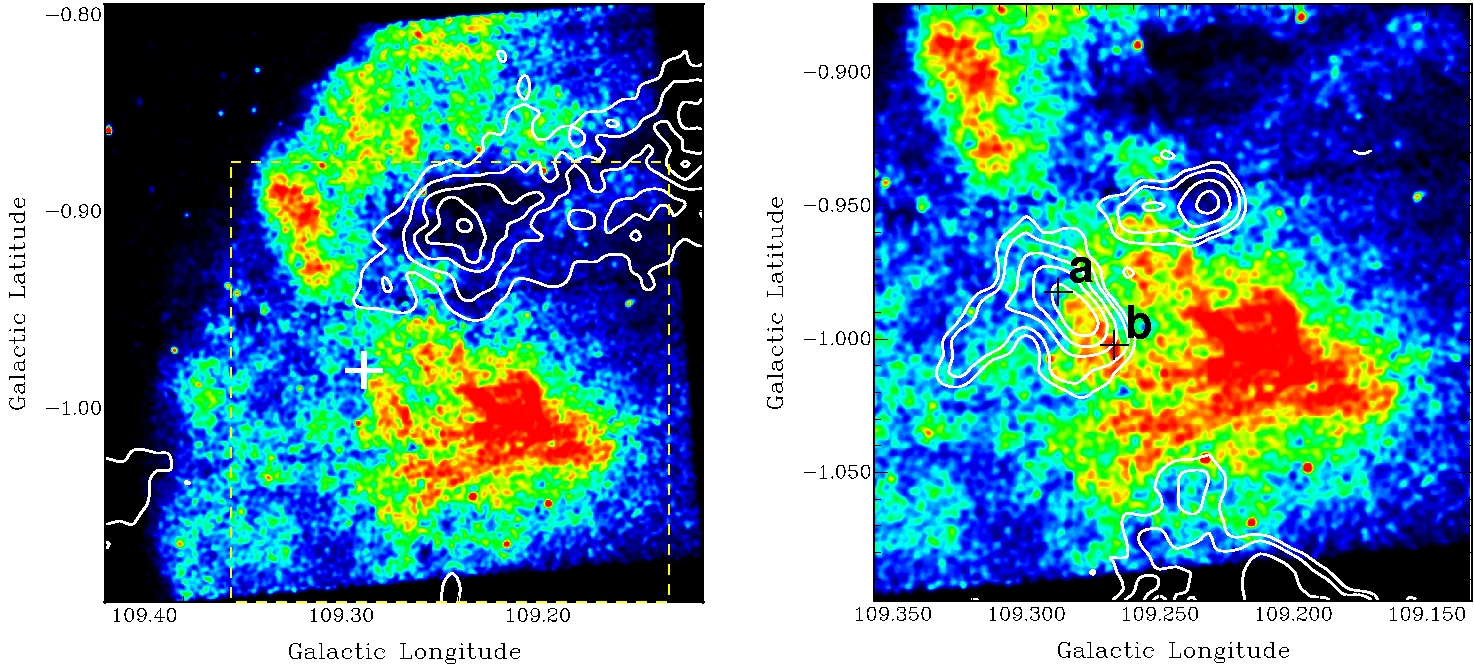.}
The distribution of molecular gas in the vicinity of the X-ray
Lobe as seen in the light of the $^{12}$CO($1-0$) line (white contours) 
overlaid on a \chandra\ image of the X-ray emission between 0.35 and 10.0~keV. 
({\it Left}) The CO data are averaged over the velocity range from 
$-51$ to $-44.5~\mathrm{km}~\mathrm{s}^{-1}$ 
and show the absorbing CO in the foreground. Contours levels are 
at 0.5, 1.0, 1.5, 2.0, and 2.5~K. 
The white cross marks the position of the IR source IRAS\,23004+5841.
The yellow dashed box shows the extent of the region shown in the right figure.
({\it Right}) The CO data are averaged over the velocity range from 
$-57$ to $-52~\mathrm{km}~\mathrm{s}^{-1}$. 
Contours levels are at 0.15, 0.25, 0.5, 0.8, and 1.2~K. Three 
clouds are seen Galactic north (up), east (left), and south (down) of the 
Lobe. The positions at which the velocity profiles in Figure \ref{covelocities} 
are taken are marked with crosses.
}
\end{figure}

In Figure \ref{chandraco} we display the distribution of molecular gas in the 
vicinity of the X-ray Lobe. In the left image, the CO 
arm discovered by \citet{1987A&A...184..279T} is shown. The anti-correlation 
of the CO emission with the \chandra\ image nicely demonstrates that this 
molecular cloud is located in the foreground and absorbs the X-ray emission 
from \ctb\ coming from behind it. In the right panel we averaged the CO 
emission over a velocity range  
more negative than that of 
the CO arm. We can 
identify three small molecular clouds surrounding the eastern part of \ctb's 
X-ray Lobe with the brightest to the 
Galactic east (on the left side of the Lobe in Fig.\,\ref{chandraco}, 
hereafter east)
, a fainter one to the 
Galactic north (above the Lobe, hereafter north)
, and 
another one in the 
Galactic south (below the Lobe, hereafter south)
which is not fully covered by our observations. 
The noise level in the image is \about40~mK.
Assuming the progenitor star exploded at or close to the current position of 
the anomalous X-ray pulsar 1E\,2259+586, the location of these clouds is
suggestive of
an interaction of the SNR shock
wave with those molecular clouds resulting in the X-ray Lobe.

Most of the molecular clouds are rather faint in the $^{13}$CO line,
which is why we cannot perform a detailed comparison with the $^{12}$CO
measurements. However, we can estimate an average brightness ratio
for each of the clouds (see Table \ref{cotab}). The 
value for the southern cloud is a bit difficult to interpret since it is not
fully covered by our observations and it seems to consist of a number
of small clouds. We find brightness ratios between 3.5 in the dense
part of the CO arm and 13 in the southern clouds. According to
\citet{1990ApJ...357..477L}, at the galactocentric radius of \ctb\ the
$^{12}$C to $^{13}$C isotope ratio $r$ should be about 63. This indicates
that we miss some of the $^{12}$CO emission and this line is optically thick.

In the following we assume local thermodynamic equilibrium and the same
excitation temperature for both isotopic species and all molecules along
the line of sight in each cloud. We determine an average optical depth
for each cloud in both lines by the following procedure:
We use the $^{12}$CO to $^{13}$CO ratio for each cloud to determine how
much $^{12}$CO emission we are missing by assuming the $^{13}$CO line is
optically thin. This can be translated to a first iteration for the optical
depth $\tau_{12}$. If both species have the same excitation temperature the
$^{12}$CO to $^{13}$CO brightness temperature ratio $r$ can be written as:
$r = (1 - e^{-\tau_{12}})/(1 - e^{-\tau_{13}})$.
From this we determine a first iteration for the optical depth $\tau_{13}$ of
the $^{13}$CO line. This is again used to determine a better value for
the missing $^{12}$CO emission and so on. This iterative procedure converges 
usually after just a few iterations. The results for $\tau_{12}$ and 
$\tau_{13}$ are listed in Table \ref{cotab}. To integrate the $^{13}$CO column 
density we actually use the $^{12}$CO data scaled to $^{13}$CO by the 
brightness ratios for each individual cloud since the signal to noise ratio is 
higher in our $^{12}$CO data. The $^{13}$CO column density
is then scaled by $5\times 10^{5}$ to determine the column density of
the \htwo\ molecules \nhtwo\ \citep{1978ApJS...37..407D}. We also estimate
H$_2$ number densities and masses of the clouds (Table \ref{cotab}). 
While the northern cloud, the southern cloud, and the faint eastern tail of the
eastern cloud have comparable \htwo\ column densities ($< 10^{20}$~cm$^{-2}$), 
the bright part of the eastern cloud that overlaps the Lobe has a higher 
\nhtwo\ $\approx 1.9 \times 10^{20}$~cm$^{-2}$.
It is interesting to note that we calculated a peak H$_2$ 
column density of $2.2 \times 10^{21}$~cm$^{-2}$ for the CO arm, which compares 
nicely with the value of $2.0 \times 10^{21}$~cm$^{-2}$ determined by
\citet{1987A&A...184..279T}.

In order to compare with these results, we extract spectra from the \chandra\ 
data in regions corresponding to the CO clouds and derive the atomic H column 
density \nh\ in the foreground by fitting the spectrum with a model including 
a thermal non-equilibrium ionization model and a foreground absorption.
For the northern cloud, we obtain \nh = 6.3 
(5.6 -- 7.0) $\times 10^{21}$~cm$^{-2}$ (90\% confidence range in parentheses).
For the larger eastern cloud, the foreground absorption of the part inside the 
Lobe is \nh\ = 4.5 (4.2 -- 4.9) $\times 10^{21}$~cm$^{-2}$, 
outside the Lobe, we get \nh\ = 5.9 (5.5 -- 6.4) $\times 10^{21}$~cm$^{-2}$,
and in the eastern tail the foreground absorption is
\nh\ = 5.0 (4.2 -- 5.4) $\times 10^{21}$~cm$^{-2}$. Although the \nhtwo\ column 
density is largest in the bright part of the eastern cloud, the
foreground \nh\ is the lowest. Therefore, the eastern cloud is not located in 
front of the Lobe.
The foreground absorption in the region of the southern cloud
is \nh\ = 6.8 (5.8 -- 7.3) $\times 10^{21}$~cm$^{-2}$. The X-ray absorption is
significantly higher in the regions corresponding to the northern cloud and the 
southern cloud than in the eastern tail of the eastern cloud. It seems that 
these two clouds are located in front of the remnant and absorb some of the 
X-ray emission.

\begin{figure}
\centering
\includegraphics[width=0.4\textwidth]{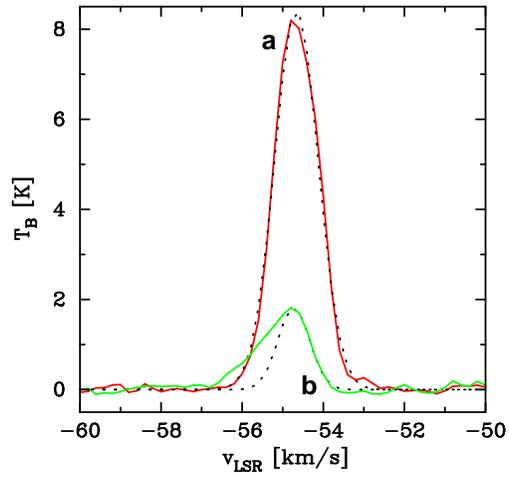}
\caption{\label{covelocities}
Velocity profile of the eastern CO cloud at the position of IRAS\,23004+5841 
(a) and in the Lobe (b). Solid lines show the data, dotted lines the Gaussian 
fits.
}
\end{figure}

As the SNR is believed to be located next to the GMC, we assume that both have 
a systematic velocity of $-51~\mathrm{km}~\mathrm{s}^{-1}$ 
\citep{1987A&A...184..279T,2002ApJ...576..169K}.
The three clouds (radial velocities between $-52$ and 
$-57~\mathrm{km}~\mathrm{s}^{-1}$) are slightly blue-shifted from the 
GMC to the west, indicating that these clouds are moving towards us 
relative to the GMC complex. 
As the bright eastern cloud seems to be related to the X-ray Lobe, we study
the velocity profiles of the $^{12}$CO emission in different parts of the 
eastern cloud. 
This cloud contains the infrared (IR) source
IRAS\,23004+5841, which has IR colors of a star forming region according to 
\citet{1989A&AS...80..149W}. 
Figure \ref{covelocities} compares the profiles taken at the center of 
IRAS\,23004+5841 (position a) and in the interior of the Lobe (position b, 
as marked in the right panel in Fig.\,\ref{chandraco}).
While the first appears Gaussian, the latter has an additional component 
towards higher negative velocities. 
The asymmetry observed in the spectrum suggests that the material has been
accelerated by the shock wave of the SNR which traveled into the cloud.
The CO line profile is only broadened by a few km~s$^{-1}$ which
indicates that the acceleration is mostly perpendicular to
the line of sight.

The estimated mass of the part of the cloud with the high velocity wing is 3 --
4~$M_{\sun}$.  
We have also taken profiles from parts of the east cloud that do not overlap
with the Lobe. The velocity profiles of the northern end of the east cloud and
the faint tail in the east show that there is a velocity gradient. 
The center of the profile changes from $-54.8~\mathrm{km}~\mathrm{s}^{-1}$ to 
$-53.5~\mathrm{km}~\mathrm{s}^{-1}$ 
with increasing distance to the Lobe, i.e.\ the eastern part of the cloud is 
red-shifted relative to the western part. This gradient indicates an
acceleration of the gas in the faint tail away from the eastern cloud. As the
gradient starts at the position of IRAS\,23004+5841, it might be an outflow
from the star forming region. 

\subsection{The Shocked Cloud}

The column densities of \htwo\ and the X-ray absorbing hydrogen indicate
that the northern cloud and the southern cloud are located in front of the 
X-ray emission. The eastern cloud, however, seems to be linked to the Lobe. 
Moreover, the CO velocity profile shows an additional blue-shifted component in 
the eastern cloud where it overlaps the Lobe, suggesting that the cloud has 
been hit by a shock. The eastern tail of the eastern cloud doesn't 
show such an additional velocity component and seems to be red-shifted
relative to the interacting part of the cloud. 

\begin{figure}
\centering
\includegraphics[width=0.4\textwidth]{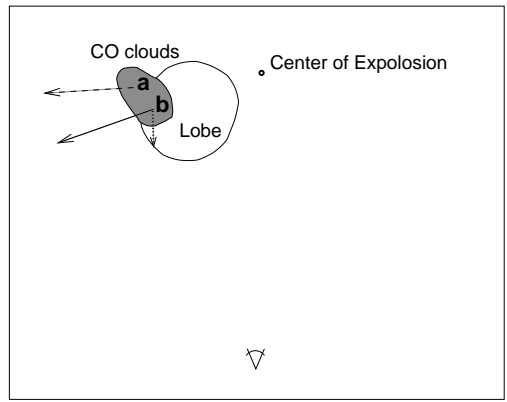}
\caption{\label{skizze}
Schematic view of \ctb\ showing the Lobe and the eastern cloud. The solid
arrow shows the directions to which the part of the cloud at position b has 
been accelerated. The velocity component directed towards us is shown with a 
dotted line. As it is not certain how the cloud component at position a is 
moving in reality, the possible movement of position a is shown with a dashed
arrow.
}
\end{figure}

Figure \ref{skizze} illustrates how the bright eastern CO cloud and the Lobe
are possibly located within the remnant. As in such a configuration, the cloud 
would have a significant velocity component towards us at position b whereas 
the acceleration is directed perpendicular to the line of sight at position a, 
a high velocity wing in the velocity profile of the cloud is only observed at 
position b. 

As we believe that the X-ray Lobe was formed by evaporation of a cloud, we 
estimate the cloud mass from the X-ray emission. 
We assume that the emission is coming only from the Lobe and that the 
evaporated cloud now fills a sphere with a radius of 3\arcmin. The XSPEC model
VNEI that we use for the spectral fits, gives the normalization $K =
[10^{-14}/ (4 \pi D^{2})] \times \int n_e n_H dV$.
The mean $K$ per arcsec$^{2}$ in the Lobe is 
2.0$\times 10^{-7}$~cm$^{-5}$~arcsec$^{-2}$. For a distance $D$ = 3~kpc
\citep{2002ApJ...576..169K} and $n_{\rm e}$ = 1.2 $n_{\rm H}$, we get 
$n_{\rm H}$ = 0.9~cm$^{-3}$ and a mass of $M = 5~M_{\sun}$ for the X-ray gas 
of the Lobe. The mass of the observed CO clouds are higher than the estimated
mass of the evaporated cloud. This could be the reason why the clouds still 
exist.

\section{Summary}

We performed observations of $^{12}$CO and $^{13}$CO as well as a \chandra\ 
observation of the region around the X-ray Lobe of \ctb.
We have discovered three CO clouds around the Lobe. All three clouds are
blue-shifted relative to the GMC in the west of \ctb. The foreground \nh\ 
indicates that two clouds in the north and in the south of the Lobe are 
located in front of the bright X-ray Lobe, whereas the east cloud might be 
connected to the Lobe. Therefore, the east CO cloud and the X-ray Lobe seem to
be evidence for an interaction between the SNR shock wave and a dense cloud.
The velocity profiles of the $^{12}$CO emission in the east cloud show that 
there is a velocity gradient in the faint tail in the east, indicating that
the eastern part of the cloud is red-shifted. The bright western part
of the cloud overlaps with the X-ray Lobe. The $^{12}$CO velocity profile at 
this position has an negative velocity wing indicating an additional 
acceleration in this part of the cloud. At this position where the CO and
the bright X-ray emission overlap, there is also an extended IR source
(IRAS\,23004+5841) which might be emission from a star forming 
region. From the new CO and X-ray data we conclude that we have
found 
strong evidence for
a shock-cloud interaction at the north-east end of the X-ray Lobe.

\acknowledgments

Support for this work was provided by the National Aeronautics and 
Space Administration through \chandra\ Award Number G04-5068X issued 
by the \chandra\ X-ray Observatory Center, which is operated by the 
Smithsonian Astrophysical Observatory for and on behalf of the 
National Aeronautics Space Administration under contract NAS8-03060.
The Dominion Radio Astrophysical Observatory is a National
Facility operated by the National Research Council of Canada.
The Five College Radio Astronomy Observatory is supported by NSF grant 
AST 01-00793.


\clearpage

\clearpage

\clearpage

\clearpage

\begin{table}
\begin{center}
\caption{\label{cotab} Calculated Cloud Characteristics and the Foreground Hydrogen Column Density}
\vspace{2mm}
\begin{tabular}{ccccccc}
\tableline\tableline\noalign{\smallskip} 
Clouds & North & East & Eastern Tail & South & CO arm & Lobe \\
\noalign{\smallskip}\tableline\noalign{\smallskip}  
 $r$ & 8.1 & 7.6 & 12.0 & 13.0 & 3.5 & ...\\
 $\tau_{12}$ & 2.2 & 2.2 & 1.7 & 1.6 & 3.2 & ...\\
 $\tau_{13}$ & 0.13 & 0.14 & 0.076 & 0.068 & 0.47 & ...\\
 \nhtwo\ [cm$^{-2}$] & 7.0e19 & 1.9e20 & 2.9e19 & 3.4e19 & 1.3e21  & ...\\
 $n_\mathrm{H_2}$ [cm$^{-3}$] & 10 & 40 & 5 & 2 & 100  & ...\\
 Mass [M$_\sun$] & 13 & 50 & 3 & 25 & 350  & ...\\
 \nh\ [cm$^{-2}$] & 6.3e21 & 5.9e21 & 5.0e21 & 6.8e21 & ... & 4.5e21 \\
\noalign{\smallskip}\tableline
\end{tabular}
\end{center}
\end{table}

\end{document}